\documentclass[conference]{IEEEtran}
\usepackage{amsmath,amssymb,amsfonts}
\usepackage[square,numbers]{natbib}
\usepackage{algorithmic}
\usepackage{graphicx}
\usepackage{textcomp}
\usepackage{xcolor}
\usepackage{verbatim}
\usepackage{multicol, blindtext}
\usepackage{wrapfig,lipsum,booktabs}
\usepackage{pifont}
\usepackage[printonlyused,nolist,nohyperlinks]{acronym}
\usepackage{fourier} 
\usepackage{array}
\usepackage{makecell}
\usepackage{multirow}

\def\BibTeX{{\rm B\kern-.05em{\sc i\kern-.025em b}\kern-.08em
    T\kern-.1667em\lower.7ex\hbox{E}\kern-.125emX}}
    
\begin{document}

\title{TrustVault: A privacy-first data wallet for the European Blockchain Services Infrastructure}

\author{\IEEEauthorblockN{Sharif Jacobino and Johan Pouwelse}
\IEEEauthorblockA{
% \textit{Department of Software Technology} \\
% \textit{Distributed Systems}\\
% \textit{Faculty of Electrical Engineering, Mathematics \& Computer Science}\\
% \textit{Delft University of Technology}
\textit{Distributed Systems, Delft University of Technology}\\
\textit{August 2022}
}
% \and
% \IEEEauthorblockN{Johan Pouwelse}
% \IEEEauthorblockA{\textit{Department of Software Technology} \\
% \textit{Distributed Systems}\\
% \textit{Faculty of Electrical Engineering, Mathematics \& Computer Science}\\
% \textit{Delft University of Technology}
% }
}

\newenvironment{todo}{
\color{red}
}

\newcommand{\RNum}[1]{\uppercase\expandafter{\romannumeral #1\relax}}

\maketitle
\thispagestyle{plain}
\pagestyle{plain}

\textbf{\emph{Abstract---}The European Union is on course to introduce a European Digital Identity that will be available to all EU citizens and businesses. This will have a huge impact on how citizens and businesses interact online. Big Tech companies currently dictate how digital identities are used. As a result, they have amassed vast amounts of private user data. Movements like Self-Sovereign Identity aim to give users control over their online identity. TrustVault is the first data wallet that gives users back
control of their identity and all their data. TrustVault allows users to store all their data on their smartphones and control with whom they share it. The user has fine-grained access control based on verifiable user attributes. EBSI connects TrustVault to the European Self-Sovereign Identity Framework allowing users to use Verifiable Credentials from public and private institutions in their access control policies. The system is serverless and has no Trusted Third Parties. TrustVault replaces the for-profit infrastructure of Big Tech with a public and transparent platform for innovation.}

% \newacronym{ssi}{SSI}{Self-Sovereign Identity}
% \newacronym{eu}{EU}{European Union}
% \newacronym{ec}{EC}{European Commission}
% \newacronym{ebsi}{EBSI}{European Blockchain Services Infrastructure}

% \newacronym{vc}{VC}{Verifiable Credential}
% \newacronym{vp}{VP}{Verifiable Presentation}

\begin{acronym}
\acro{SSI}{Self-Sovereign Identity}
\acro{EU}{European Union}
\acro{EC}{European Commission}
\acro{EBSI}{European Blockchain Services Infrastructure}
\acro{ESSIF}{European Self-Sovereign-Identity Framework}

\acro{IDP}{identity provider}

\acro{VC}{Verifiable Credential}
\acro{VP}{Verifiable Presentation}

\acro{ABAC}{Attribute-Based Access Control}
\acro{ABE}{Attribute-Based Encryption}
% \acro{AC}{access control}
% \acro{AP}{access policy}

\acro{VDR}{Verifiable Data Registry}
\acro{TIR}{Trusted Issuers Registry}
\acro{TSR}{Trusted Schemas Registry}
\acro{SIC}{Self-issued Credential}
\acro{DID}{decentralised identity}

\acro{GDPR}{General Data Protection Regulation}

\acro{EC}{Elliptic Curve}

\acro{P2P}{peer-to-peer}
% \acro{DV}{data vault}
% \acro{DW}{digital wallet}
\acro{TCID}{TrustChain IDentity}
\acro{ZKP}{zero-knowledge proof}
\acro{TTP}{Trusted Third Party}

\acro{UI}{user interface}

\acro{UDP}{User Datagram Protocol}
\acro{JWT}{JSON Web Token}

\acro{DPV}{Data Privacy Vault}
\end{acronym}

\section{Introduction}
Internet users have little control over where and how their data is stored and used online. Big Tech companies store gigabytes of data about you and know exactly which online services you use \cite{are-you-ready}. User data is a precious asset and the primary source of income for such companies. Billions of people rely on Big Tech monopolies to store their data and voluntarily give up control and ownership over it. Much of this data is deeply personal and valuable, such as intimate photos of our friends and family. Public and policy trust in Big Tech has been breaking down in recent years (also called the "techlash") following major scandals, rampant misinformation campaigns, and a perceived consolidation of power \cite{birch2021data}. Nearly five decades after the invention of public key cryptography, we still lack a good solution for people to manage their digital identity and efficiently share encrypted data directly with each other, certainly at a massive scale. Various movements aim to halter Big Tech's power and give back control to the users. These movements are powered by technologies like blockchain and Self-Sovereign Identity (SSI), which promise to improve how we interact with online services and each other. Distributed computing has progressed to a point where a truly distributed identity system, where trust is diffused and not under the control of any entity, is possible.

\acl{SSI}, sometimes referred to as "The Internet's missing identity layer" is an attempt at satisfying the following requirements for a digital identity \cite{tobin2016inevitable}:
\begin{itemize}
    \item Security: protecting identity information from unintentional disclosure.
    \item Control: the identity owner determines who can access their data and under what circumstances
    \item Portability: user identity must not be tied to a single service or provider
\end{itemize}
These properties are what make \ac{SSI} a tool that will inevitably shift power away from centralised organisations and towards the people.

The \ac{EU} is not unaware of these movements and is ramping up its efforts for bringing transformation into the digital sphere with projects such as Europe's Digital Decade \cite{eu-digital-decade}. In September 2020, the president of the \ac{EU} declared that a European Digital Identity would be made available to all \ac{EU} citizens and they all will be able to have a digital wallet \cite{eu-digital-identity}.

\begin{quote}
    \emph{"Every time an App or website asks us to create a new digital identity or to easily log on via a big platform, we have no idea what happens to our data in reality."} Ursula von der Leyen, President of the European Commission\\
\end{quote}

One of the goals of the \ac{EU} is to improve the way citizens, businesses and public administrations share information and trust each other, and simplify verification processes for cross-border services using blockchain technology \cite{ebsi-home}. Its proposed solution to reduce our reliance on Big Tech is the \ac{EBSI}. As at May 2022, there was €57 million in funding for large scale trials\footnote{https://ec.europa.eu/digital-building-blocks/wikis/display/EBSI/EBSI+Grants} and in June 2022 a tender valued at €26 for the development of the European Digital Identity Wallet million was published by the EU \footnote{https://ted.europa.eu/udl?uri=TED:NOTICE:309685-2022:TEXT:EN:HTML\&tabId=1}. \ac{EBSI} uses \acl{SSI} to reduce the time and cost of verifying the authenticity of documents and information shared on the \ac{EBSI} network. \ac{EU} citizens will be able to download a wallet from the app store and interact with \ac{EBSI}\footnote{https://www.thalesgroup.com/en/worldwide-digital-identity-and-security/government/magazine/eu-digital-id-wallet-coming-heres-what}. Wide-scale adoption will have a significant impact on the digital lives of \ac{EU} citizens.

While \ac{EBSI} and \ac{SSI} in general can make users sovereign over their identity, non-identity data remains on the servers of centralised applications, unable to be used within other applications. If you have had enough of Facebook, migrating your photos to another photo-sharing app would be a huge undertaking. It would also be near impossible to completely control who has access to your data on a remote server.

% This work aims to answer one question: \emph{How can we give user back control of their own identity and their data?} We want to accelerate the European identity and data self-sovereignty movement by providing a EBSI-certified data wallet with advance data sharing capabilities. 

This work aims to solve these problems by developing a data wallet with advanced data sharing capabilities that leverage \ac{SSI} to provide users with true sovereignty over their data. The contribution of this work is TrustVault: A privacy-first data wallet deployed on the TrustChain Super App. TrustVault consists of a secure data vault and an EBSI conformant digital wallet. The data vault stores the user's data locally and provides fine-grained access control for the stored files. The digital wallet holds \acp{VC} obtained from the \ac{EBSI} network and presents these credentials to peers using TrustVault. \acp{VC} contain attribute claims that function as access tokens to other users' data vaults. Using \acp{VC} as a basis for \acl{ABAC} for personal data storage is a novel concept that extends the notion of self-sovereignty over personal identity to personal data. This base implementation lets you browse through photos of your peers and demonstrates TrustVault's ability to be used for zero-server applications. Users connect directly to your TrustVault and their credentials are automatically matched against your predefined access policies. Users only see the photos that you allow them to see. Our openness-by-design ecosystem encourages permissionless innovation and competition. Anyone is able to develop new decentralised applications that can interact with data in your TrustVault.\\

This research contributes (\RNum{1}) a GDPR-compliant by design data wallet that (\RNum{2}) integrates two \acl{SSI} frameworks with a novel use of \aclp{VC} for access control in a \acl{P2P} setting (\RNum{3}), creating a societal infrastructure for trusted data sharing.

\section{Problem description}\label{sec:problem-description}
The goal of this study is to design a system that gives users sovereignty not just over their identity, but also over their data. In other words, can we extend the security, control and portability properties of \ac{SSI} from identity to data in general? The system has to be a part of the critical societal infrastructure being developed by the \ac{EU} to reduce reliance on Big Tech.
Web applications that see a lot of user data are prime targets for hackers \cite{musa2015systematic}. The reward for disrupting important services and stealing confidential data is huge. Much effort goes into securing centralised applications with frequent penetration testing, better software development methods, hardening techniques like encryption and so on. Yet, even if user data is encrypted, much information can be inferred from the large amount of metadata collected by web applications with statistical analysis, possibly breaching user privacy \cite{safebook}. Dispersing data throughout a network lowers the risk of large-scale data breaches and makes the system more fault tolerant. As long as your data is on a remote server, it is not truly under your full control. Soft access control is hard to enforce if parties can be malicious \cite{ac-enforce}. Hard access control (enforced with cryptography) is either not very flexible when using public-key cryptography or introduces Trusted Third Parties (TTP) in the case of most \acl{ABE} schemes \cite{muller2008distributed}. Most importantly, even systems that offer fine-grained access control without TTPs like distributed \acl{ABE} schemes do not prevent censorship \cite{lee2021dq} by centralised applications. Data portability is a personal right established in the \ac{GDPR} \cite{right-portability}. This is in direct conflict with the desire of companies to retain users and their data. Data is often tightly coupled to the application, complicating transitioning data between services. Regulations and public pressure is forcing companies to adopt or support standard formats. Data must still to be exported from one application and imported into the other. This step can be simplified or even eliminated.

A system where users have true sovereignty over data has to have the following properties: 
\begin{itemize}
    \item Data storage has to be decentralised on devices under the control of the data owner.
    \item Access control has to be verifiably authenticated, fine-grained and resolutely enforced.
    \item Data has to be decoupled from applications.
\end{itemize}

Applications access user data at the discretion of the user. Certain applications require users to access data on another user's device. The requesting user has to satisfy the access policy set in place by the host user for the desired data. Secure access control requires a user’s authentication to be verified before enforcement \cite{sandhu1994access}. \ac{SSI} solves this problem in a way that keeps users in control of their identity. Actually, SSI makes it possible for any attribute of a user to be verifiable through \acp{VC}. Access policies can be defined in a fine-grained manner for arbitrary verifiable attributes.

\ac{EBSI} will be the connecting piece to the societal infrastructure for identity once in production. EU citizens will have credentials from public and private institutions such as driver's licence, diplomas and club membership in digital form. These can all be used to enable the automatic sharing of data between EU citizens based on these credentials. In section \ref{sec:background} we elaborate on concepts relevant to this work. In section \ref{sec:architecture-design} the architecture and design of TrustVault are presented and in section \ref{sec:implementation-evaluation} we discuss the implementation and evaluate the system. In section \ref{sec:related-work} we go over related work and we end with a conclusion and future work proposals in section \ref{sec:conclussion-fw}.
\section{Background}\label{sec:background}
\subsection{Self-Sovereign Identity}
\ac{SSI} is a decentralised model of digital identity developed to address the shortcomings of the previous internet identity models \cite{preukschat2021self}. With centralised identities, centralised institutions such as governments and banks issue credentials that allow citizens to interact with services and each other. On the internet you would establish an account with every website, service or application. In this model, all the data about you belongs to the issuing party, can't be reused, and is out of your control.\\
The federated identity model introduces \acp{IDP}. \acp{IDP} allow you to have one account that can be used to interact with any service that supports that \ac{IDP}. This is the mechanism behind the social login buttons (Login with Facebook) widely found on the internet today. Federated identity simplified managing accounts for every service to managing a few accounts at a few \acp{IDP}. All our identity data and information about when or how we use our federated identities is now concentrated in these Tech Giants, raising many privacy concerns.\\
The rise of blockchain technology inspired the decentralised identity model. This model is not based on accounts with centralised institutions or \acp{IDP} but on direct relationships between peers. No party controls or owns the relationship. Users are in complete control of their identity data, how it is shared, and with whom. Peers establish private connections by securely exchanging public keys whereby blockchains serve as decentralised public key infrastructures. This model closest resembles how we manage our identities in the real world: with wallets containing credentials obtained from trusted parties which can be shown to other parties to initiate an interaction.
There are several deployed \ac{DID} frameworks built on top of ledgers purpose-built for decentralised identity like Sovrin \cite{sovrin} (based on the Hyperledger Indy framework\cite{hl-indy}) and ledgers repurposed for \ac{SSI}, such as  \cite{deployment-bc-ssi-stokkink} (using TrustChain \cite{otte2020trustchain}) and Ethereum \cite{eth-did}.

\begin{figure}
\centering
\includegraphics[width=0.45\textwidth]{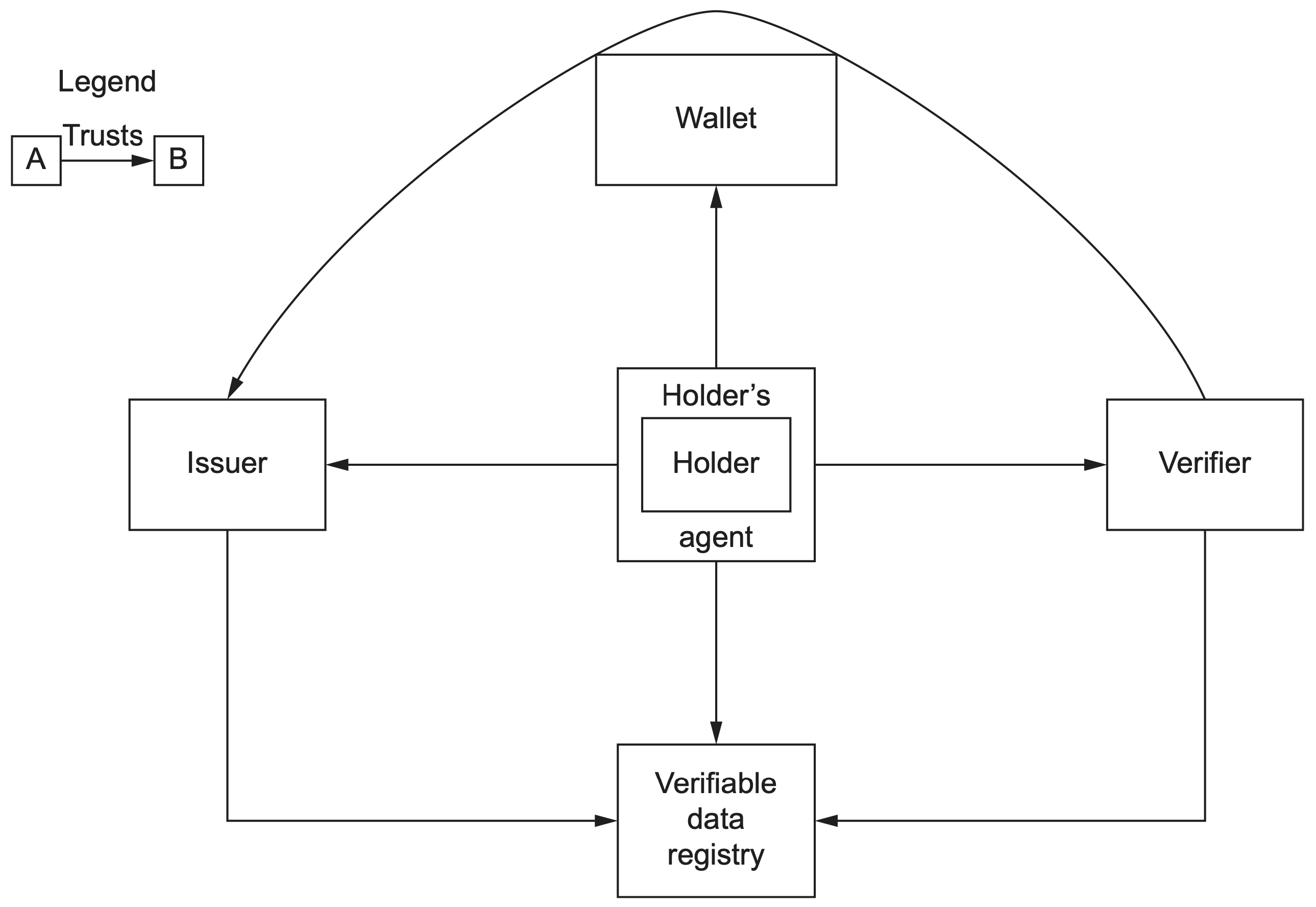}
\caption{VC trust model \cite{preukschat2021self}.}
\label{fig:vc-trust-model}
\end{figure}

Verifiable Credentials are the building blocks of \ac{SSI}. Much like physical credentials, \acp{VC} contain claims about your identity that some authority claims are true about you. You can then use this \ac{VC} to convince others who trust said authority of the validity of these claims. The trust relationship between issuers, holders/provers, and verifiers is shown in Figure \ref{fig:vc-trust-model}. Issuers put digital signatures on credentials that are cryptographically verifiable. They are trusted to issue true credentials and be authoritative on the attributes they attest to. Verifiers request proof about identity claims they need to be convinced of. They do not need to have any direct relationship with issuers. They just need to trust an issuer's ability to make correct assertions. Holders ultimately have the choice to respond to a request with a \ac{VP}: a \ac{VC} with a digital signature of the prover. Holders trust verifiers to keep their credentials confidential. The \ac{VDR}, where \acp{DID}, public keys and schemas are registered, must be trusted by every party to be accurate and tamper-evident. That is why public ledgers are a good fit for the function of \ac{VDR}. The holder's credentials and cryptographic keys are stored in a digital wallet. The wallet is trusted to store \acp{VC} securely. Digital agents wrap users' digital wallets and establish communication with other agents to exchange credentials.

\subsection{European Blockchain Services Infrastructure}
\acf{EBSI} is a distributed network that runs a public blockchain to host public and private services that want to leverage the benefits of blockchain technology. Their objective is to offer secure and private cross-border public services among EU member states. The main services that EBSI aims to facilitate are:
\begin{enumerate}
    \item Notarization: using the blockchain to make digital audit trails and automate compliance checks.
    \item Diplomas: giving citizens control over their educational credentials and lowering the cost of verifying documents.
    \item \acf{ESSIF}: serves as a verifiable registry and communication channel for an SSI framework across Europe.
\end{enumerate}

Relevant to this work is \ac{ESSIF}, enabling the exchange of \acp{VC} on EBSI. This service encourages European citizens to adopt SSI to improve the identity verification process with public services and private companies across European borders. The EBSI blockchain serves as the \ac{VDR} in the \ac{ESSIF} framework, where public keys of users and trusted applications can be looked up.

The EBSI architecture consists of three layers: the Infrastructure layer, the Chain and Storage layer, and the Core Service layer. The Infrastructure layer contains the elements required to set up an EBSI node and form a network. Every \ac{EU} Member State is allowed to run nodes, distributing trust over all the members. The Chain and Storage layer contains the blockchain protocols and adds off-chain storage. This is where the smart contracts for the different verifiable registries such as the DID registry, the \ac{TIR} and the \ac{TSR} are defined. These elements are segregated to make it possible to interoperate with different blockchain networks. The Core Service layer is the interface to the lower layers. It contains the API endpoints to interact with the verifiable registries and secondary services like the Notifications service.

\subsection{Attribute-Based Access Control}
\acf{ABAC} is an access control model that controls access to objects by evaluating rules against the attributes of entities \cite{hu2015attribute}. This allows for fine-grained access control because of the large set of possible combinations of attributes that can feed into an access control decision and, consequently, a large set of possible rules for policies, only limited by the richness of the available set of attributes. \ac{ABAC} makes it possible to define access control policies without prior knowledge of who will need access and no list needs to be modified to accommodate new users. Access control decisions are purely based on the presented set of attributes. An essential requirement for \ac{ABAC} is that attribute values are correctly associated with the subject. False attributes can grant unintended access to data.
\section{System Architecture and Design}\label{sec:architecture-design}
In this section we discuss how the different internal and external components come together to form the TrustVault architecture. We then go into how we integrated Verifiable Credentials into the access control mechanism to achieve fine-grained access control. We then discuss the design for a tamper-proof access log.  Finally, we explain the security measures taken to protect data in TrustVault.

\begin{figure*}
    \centering
    \includegraphics[width=0.55\linewidth]{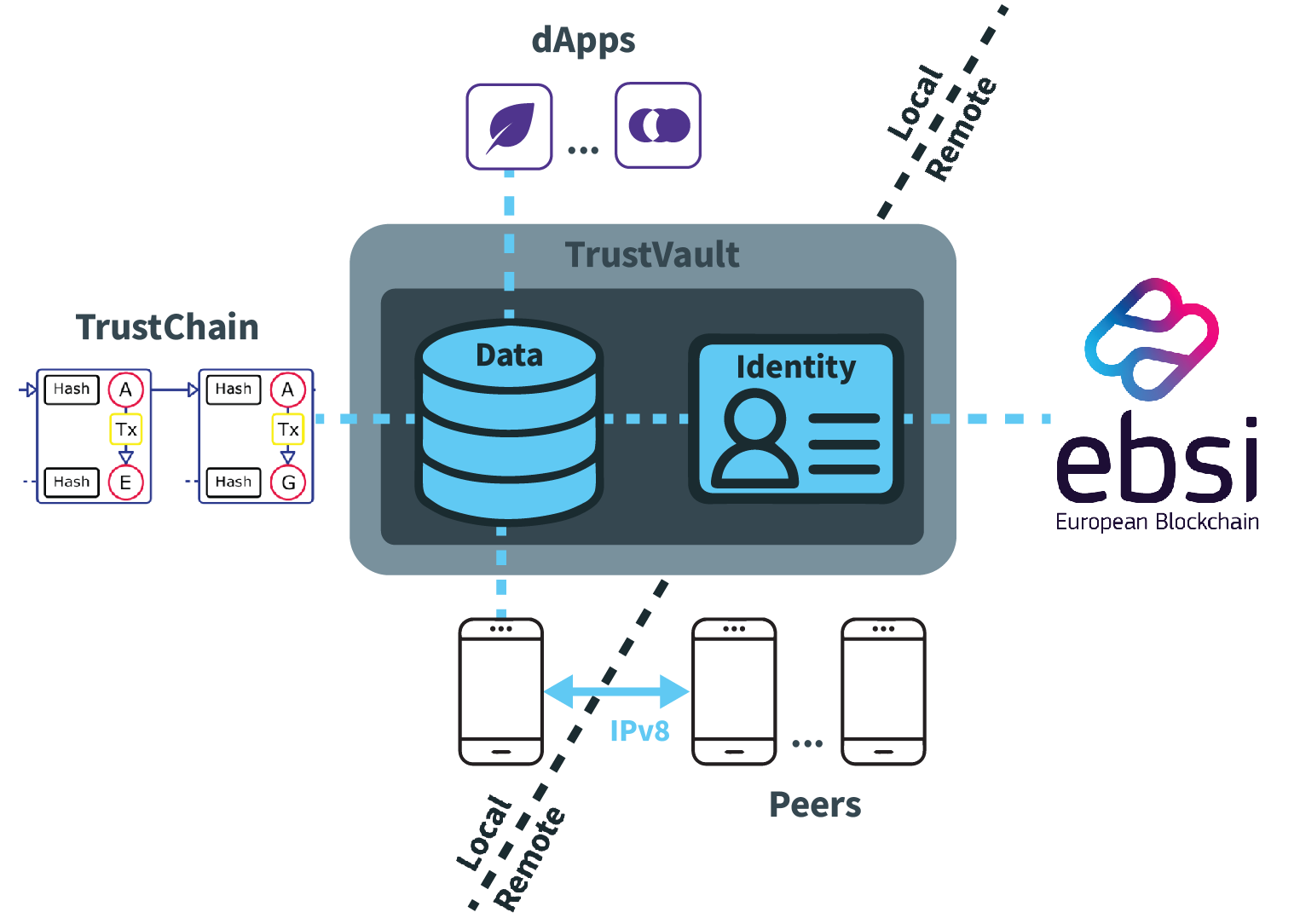}
    \caption{TrustVault Architecture.}
    \label{fig:architecture}
\end{figure*}

\subsection{Architecture}
TrustVault is a mobile agent consisting of two parts: a secure data vault and a digital wallet. The architecture of TrustVault is shown in Figure \ref{fig:architecture}. A software agent is a computer program that can act on behalf of an individual autonomously\footnote{https://www.britannica.com/technology/software-agent}. TrustVault autonomously enforces the user's access policies for the data vault and manages the credentials in the digital wallet. The data vault uses the IPv8 networking protocol for \ac{P2P} data sharing. IPv8 is a fully decentralised architecture for private and authenticated communication \cite{ipv8}. Peers communicate directly with each other without the need for servers, protecting their privacy. The protocol is built around communities that represent distinct services. Communities provide the ability for peer discovery and define service-specific messages that can be exchanged between peers. The data vault has it's own community that implements the data vault protocol. The data vault protocol is based on 5 messages: \emph{accessibleFilesRequest}, \emph{accessibleFilesResponse}, \emph{fileRequest}, \emph{fileResponse} and \emph{fileRequestFailed}. IPv8 abstracts away physical addresses and allows peers to be identified by their public keys. Connections between peers are maintained when IP addresses change, even behind NAT boxes and firewalls, by using a UDP hole-punching technique. The user can select a peer to interact with from a list of all the peers in the DataVault Community.

The data vault functions as a personal file server to the DataVault Community. The latest smartphones have storage capacities rivaling laptops. We are also used to having a large amount of personal files, mostly photos, on our smartphones. Mobile internet speeds are also approaching landline internet speed\footnote{https://www.statista.com/statistics/689876/average-mobile-speeds-download-and-upload-in-western-europe/}, especially with the rollout of 5G. The data vault stores and organises data in a closed-off directory on the phone's file system. The digital wallet also stores \acp{VC} and key material in a closed-off directory and interacts with the \ac{EBSI} Core Services Layer via its REST API. This architecture assumes \ac{EBSI} and its infrastructure has negligible downtime.

TrustVault's open architecture encourages the development of new decentralised applications that can read and write data to the user's data vault. Different applications can provide different ways of interacting with data in users' vaults. This makes for a more competitive ecosystem as user data is completely portable between applications.

TrustVault is GDPR compliant by design. All data is stored with and hosted by the user. There is no data storage or processing by third parties eliminating the need for Privacy Officers and Data Protection Officers. The system is specifically designed to not rely on cloud infrastructure.

% TrustVault is a component of the the TrustChain Super App and uses the IPv8 protocol for peer-2-peer communication. There is a specific IPv8 overlay for discovering and interacting with peers using TrustVault. TrustVault uses the Android app-specific directories to store data. Every file has a unique path down the a directory tree starting from the root directory. The user is able to create, delete and move files and folders, much like a traditional file system. The wallet is designed to support EBSI's Verifiable Credentials lifecycle \cite{ebsi-vclc}. As a holder, the wallet is capable of generating EBSI did's, store credentials and key material, request new credentials from issuers and present credentials to verifiers. As a verifier, the wallet is capable of verifying credentials and issuers.

\subsection{Access Control}\label{sec:access-management}
Files and folders, including the root folder, have an associated meta-data file that includes the file or folder's local access policy $\pi(f)$. To access a file, the file's global policy $\Pi(f)$, meaning every policy along the file's root path, must be satisfied. With $P(f)$ denoting the parent folder of $f$, $P(root) = \emptyset$ and $\Pi(\emptyset) = \emptyset$, global policies follow this recursive definition: $\Pi(f) = \pi(f) \wedge \Pi(P(f))$. Practically this means that policies are inherited from parent folders. An effective way of setting access policies is to have minimal or no restrictions on the root folder and have increasingly specific and restrictive policies for sub-folders.

An access policy is a binary boolean expression tree and the leaves are attribute rule expressions that are evaluated at access time. Attribute rules are triplets in the form of $(attribute,\ operator,\ value)$. An example policy would be $(age \geq 18 \wedge (university = TU\ Delft \vee issuer = me))$. To satisfy this policy, the prover has to present a \ac{VC} that asserts that their age is over 18, e.g. a government ID, and either a proof-of-enrolment from the TU Delft or a \ac{VC} issued by the verifier (owner of the TrustVault). The \ac{VC} is first verified by the wallet and then evaluated against the global access policy. The $age \geq 18$ rule can be satisfied with a predicate proof. A predicate proof proves a boolean statement about a value without having to reveal the value. The user interface lets the user add or remove nodes to the policy tree. Additionally, the user can define read+write policies or separate read and write policies for more granular control.

\begin{figure}
    \centering
    \includegraphics[width=0.48\textwidth]{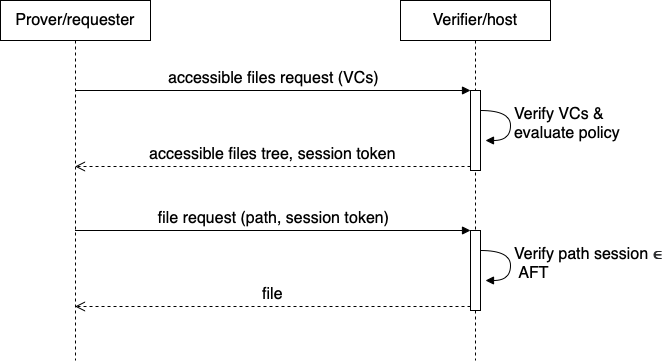}
    \caption{File exchange between TrustVaults.}
    \label{fig:file_exchange}
\end{figure}

In a protocol run, the prover is referred to as the requester and the verifier as the host. The requester first makes an \emph{accessibleFilesRequest}. An \emph{accessibleFilesRequest} must include a set of verifiable credentials as an access token. TrustVault supports credentials from two SSI frameworks: \ac{TCID} attestations and EBSI Verifiable Credentials. TCID attestations contain a single verifiable attribute claim. A fingerprint of the holder's IPv8 public key is included in the attestation to prevent replay attacks from unintended provers. The attestor's public key is attached to the attestation and does not have to be looked up in a registry. EBSI \acp{VC} follow the W3C Verifiable Credentials Data Model\footnote{https://www.w3.org/TR/vc-data-model/} standard that can contain multiple attribute claims. The prover assembles one or more VCs into a \acf{VP}. Unlike TCID attestations, W3C VCs are not coupled to a public key but to a \ac{DID}. There can be multiple key pairs associated with a DID. The prover signs the \ac{VP} with one of the private keys associated with the DID in the enclosed VCs to authenticate itself as the holder. The verifier must look up the public keys of the DID in the EBSI DID Registry to verify the authenticity of the VP. Credential access tokens with invalid signatures are not included for access policy evaluation. The principles of SSI lets the prover keep control over their identity by deciding which credentials to disclose to the verifier. Coincidentally, the verifier is given confidence that their \ac{ABAC} policies ensure that only users that authentically possess the required attributes can access their data. After policy evaluation, the host returns an \emph{accessibleFilesResponse} with a new session token that encapsulates the directory sub-tree of the file paths with global access policies satisfied by the provided credentials. Session tokens are \acp{JWT} signed by the host. Besides the directory tree, a fingerprint of the holder's IPv8 public key is also included in the session token. The signature ensures the integrity of the directory tree and the fingerprint and prevents replay attacks in the same manner as TCID attestations. The directory sub-tree is used to dynamically recreate a copy of the host's data vault on the requesting device. The actual files are retrieved on demand (\emph{fileRequest}) to prevent retrieving files that are not needed. Retrieved files are cached to avoid fetching the same files multiple times, without storing them permanently on the device. The sub-tree in the session token ensures that no files are served that are not covered by the original credentials. The verifier is freed from having to keep a mapping of session tokens to accessible files for subsequent requests and use up limited memory space. Session tokens have an expiry time and a request with an expired session token fails and the requester is notified with a \emph{fileRequestFailed} message to make a new \emph{accessiblefiles} request. This interaction is depicted in figure \ref{fig:file_exchange}.

\subsection{Self-Issued Credentials}\label{sec:self-issued-credentials}
\ac{VC} meta-data contains data unrelated to the identity of the credential subject, such as the issuer and issuance date for which access policy rules can also be made. We make use of this feature to create a new set of access policies based on \acp{SIC}. \acp{SIC} serve a similar but more expressive function than follow/friend requests in traditional social networks. The issuer can add extra attributes to a \ac{SIC} to make the context of the social connection more specific. This is particularly useful when you want to control access based on claims that a trusted issuer will not assert. For example, you have some photos you took with some people you met on holiday in Italy. You can issue a credential to them that asserts that you have met on holiday, giving them access only to the photos in your vault with the corresponding access policy. It is possible to model complex social connections in this manner, making TrustVault a well-suited data store for decentralised social applications.
\ac{EBSI} only allows \acp{VC} to be issued by trusted issuers. These are parties that have completed a verification and/or accreditation process to be registered on the \acf{TIR}. However, \acp{SIC} are only intended to be presented back to the issuer. \acp{SIC} can therefore be exchanged directly between peers, bypassing \ac{EBSI}. The issuer can verify a \ac{SIC} without having to consult the \ac{TIR}. \ac{TCID} inherently supports \acp{SIC}. In \ac{TCID} each agent has its own local list of Trusted Issuers eliminating the need to consult an external registry altogether.

\begin{comment}
Using \acp{VC} allows the host TrustVault to verify the authenticity of a credential, including verifying if the credential was issued by the controller of the vault. \acp{SIC} can serve a similar but more expressive function than friend connections in traditional social networks. Extra attributes can be added to SICs to make a distinction between different types of friendships. These distinction can be used to define even more targeted policies. The friendship token in the previous example is a self-issued credential that grants vault access to those that are given one. This makes TrustVault well-suited for social applications.
\end{comment}

\subsection{Tamper-proof access log}
Access control is completely automated without the intervention of the TrustVault owner. This makes it impossible to keep track of who has been given access to which files. This is remediated by recording \emph{accessibleFilesRequest} on TrustChain for each session. The owner sends a transaction to the requester with a bloom filter containing the accessible files from the request. A bloom filter is a randomized data structure for representing a set of elements that supports membership queries with no false negative and a small false positive probability \cite{bloom}. This forms a timestamped, tamper-proof record of the files made accessible to the requester. TrustChain transactions have to be signed by both the sender and the recipient. The requester's approval of the record is thus made irrefutable. This also guarantees the integrity of the record. In case of an audit or dispute, this record can be referenced and the bloom filter can be queried to prove with high probability that a specific file was offered to a specific user. We opted not to log each \emph{fileRequest}, which would give more credence to the access log in case of disputes, but the number of log entries would be multiple times higher, causing TrustChain to take more space on the user's phone.

\subsection{Data protection}
\begin{figure}
    % \begin{minipage}{0.49\columnwidth}
        \centering
        \includegraphics[width=0.9\columnwidth]{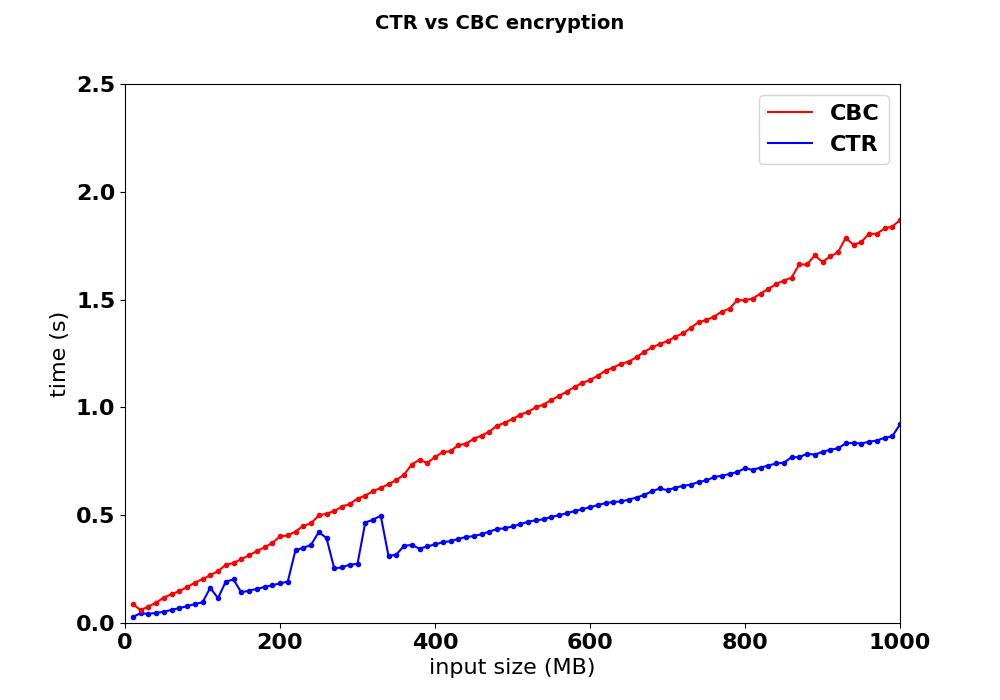}
        \caption{AES encryption CTR  vs CBC mode.}
        \label{fig:encryption-metrics}
    % \end{minipage}
    % \hfill
    % \begin{minipage}{0.49\columnwidth}
    % \end{minipage}
\end{figure}

\begin{figure}
    \centering
    \includegraphics[width=0.9\columnwidth]{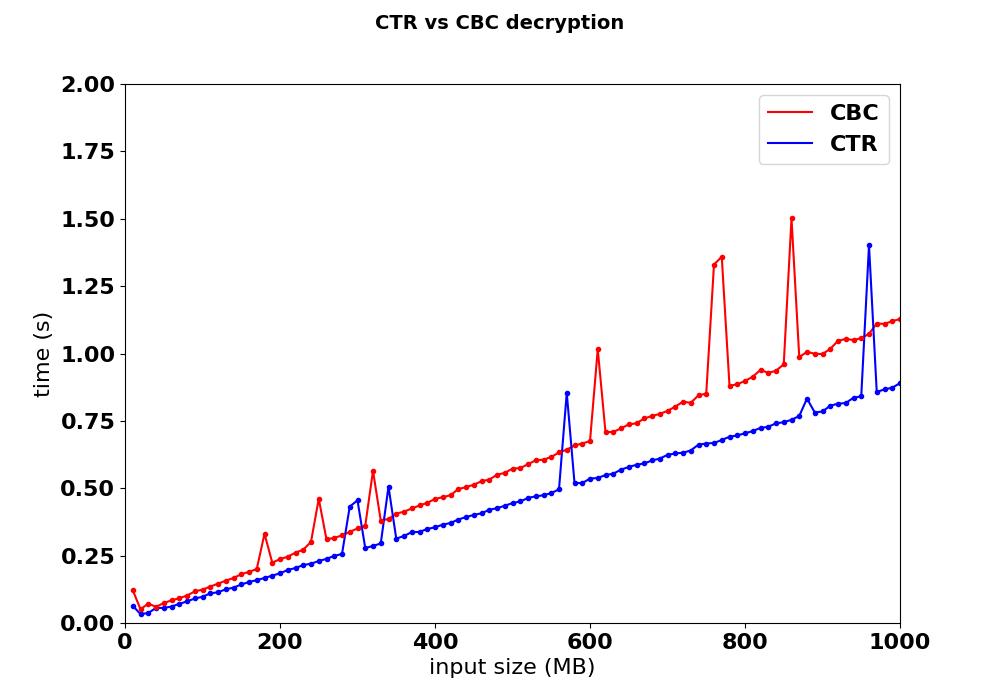}
    \caption{AES decryption CTR  vs CBC mode.}
    \label{fig:decryption-metrics}
\end{figure}

As a data wallet for \ac{EU} citizens, it is crucial that personal data and the user's right to privacy are protected in line with the \ac{GDPR}. An essential measure is to have data in the system encrypted at rest and in transit. When the TrustVault is inactive, all files are encrypted with AES in Counter mode. Counter mode is great for encrypting/decrypting large amounts of data compared to the standard Cipher Block Chaining mode because blocks can be processed in parallel. Figures \ref{fig:encryption-metrics} and \ref{fig:decryption-metrics} show the encryption and decryption of 1GB performance of a Samsung Galaxy S8. This includes \acp{VC} stored in the wallet. A password is required to "unlock" the TrustVault and "lock" it again when closing the app. The encryption key is derived from the password using PBKDF2. When transmitting files, IPv8's end-to-end encryption is used. Data packets are asymmetrically encrypted for the recipient and signed for confidentiality, integrity and authenticity of transferred files.

\section{Implementation and Evaluation}\label{sec:implementation-evaluation}
This section describes the implementation process of TrustVault and the digital wallet for EBSI specifically. We then evaluate the system's privacy protection and security. We explain our experimental setup and provide some insight into the system's performance.

\subsection{Implementation}
\begin{figure}
    \centering
    \includegraphics[width=0.48\textwidth]{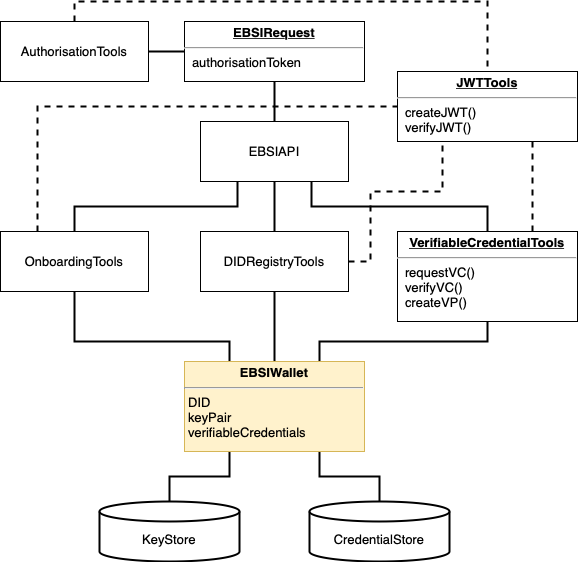}
    \caption{Digital wallet components.}
    \label{fig:wallet_components}
\end{figure}

TrustVault is made for Android and is implemented entirely in Kotlin\footnote{https://github.com/Tribler/trustchain-superapp/pull/122}. It is part of the TrustChain Super App, the collection of decentralised apps running on IPv8 and the TrustChain ledger. The codebase includes a fork of walt.id SSI kit. The open source code for SSI kit is also written in Kotlin. However, it is developed as a command line tool and does not run on Android out of the box. Changes needed to make it compatible with Android include modifications to IO operations with the file system and replacing networking and crypto libraries unavailable on Android. Figure \ref{fig:wallet_components} depicts a component diagram of the digital wallet for EBSI. The core component is the EBSIWallet class. The EBSIWallet stores and retrieves cryptographic keys and credentials from the KeyStore and CredentialStore using walt.id SSI kit key store and credential store functionalities. OnboardingTools, DIDRegistryTools and VerifiableCredentialTools implement common flows used in TrustVault like registering with EBSI, posting or updating a DID document on the DID Registry, or requesting a credential from an issuer. Upon the first launch of TrustVault, an \ac{EBSI} \ac{DID} and an \ac{EC} key pair are created and registered with \ac{EBSI}. Verifiers can then look up the user's public key in \ac{EBSI}'s DID Registry. Subsequently, the user can obtain \acp{VC} from trusted issuers on the network. These could be private or public entities all over the \ac{EU}, making the attributes for which access policy rules can be defined very diverse. As the official framework for European digital identities, EBSI makes official government credentials available for access control use. The easiest way to obtain \acp{VC} at the moment of writing is using the Conformance Test Mock Issuer. These tools share common APIs grouped together in the EBSIAPI class. Requests and responses to and from the EBSI API often included \acp{JWT}. JWTTools generates and verifies \acp{JWT} for the different tools. The DIDRegistryTools looks up public keys in the DID Registry required for JWT verification. The AuthorisationTools request short-term session tokens from the Authorisation Service using a long-term authorisation token obtained during onboarding. Session tokens are embedded in request headers when accessing protected resources or doing write operations on EBSI.

\begin{figure}
    \centering
    \includegraphics[width=0.48\textwidth]{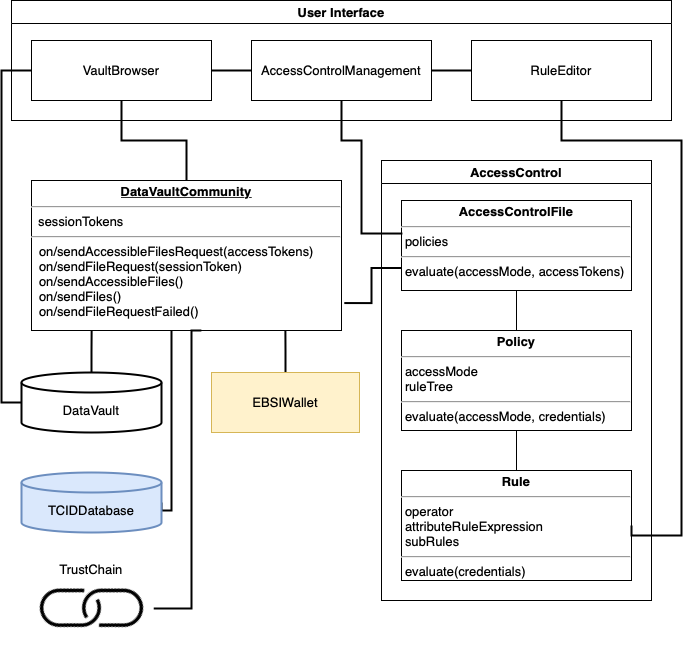}
    \caption{Data vault components.}
    \label{fig:data_vault_components}
\end{figure}

\begin{figure*}
    \begin{minipage}{0.495\columnwidth}
        \centering
        \includegraphics[width=.9\linewidth]{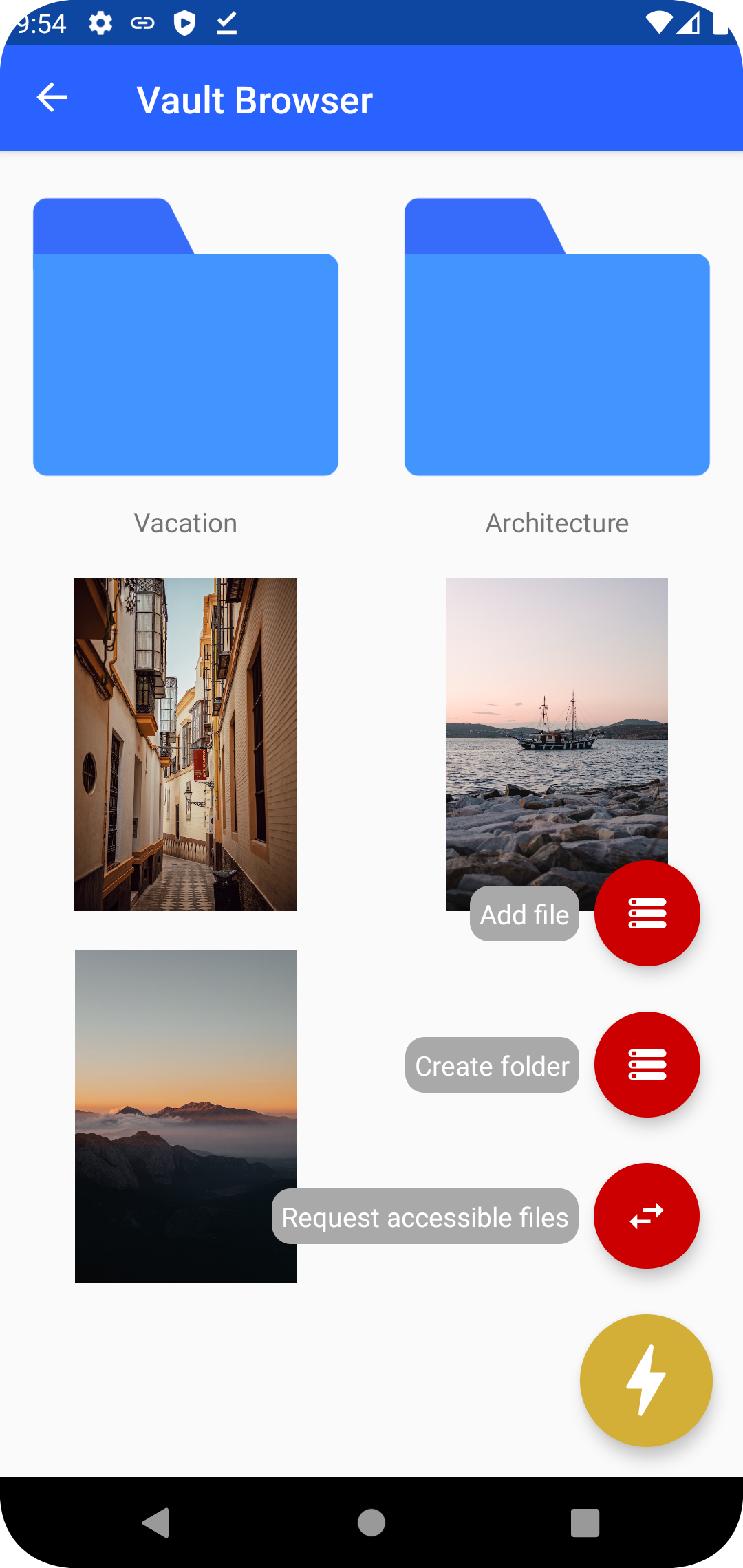}
        \caption{Data vault browser.}
        \label{fig:vault-browser}
    \end{minipage}
    \hfill
    \begin{minipage}{0.495\columnwidth}
        \centering
        \includegraphics[width=.9\linewidth]{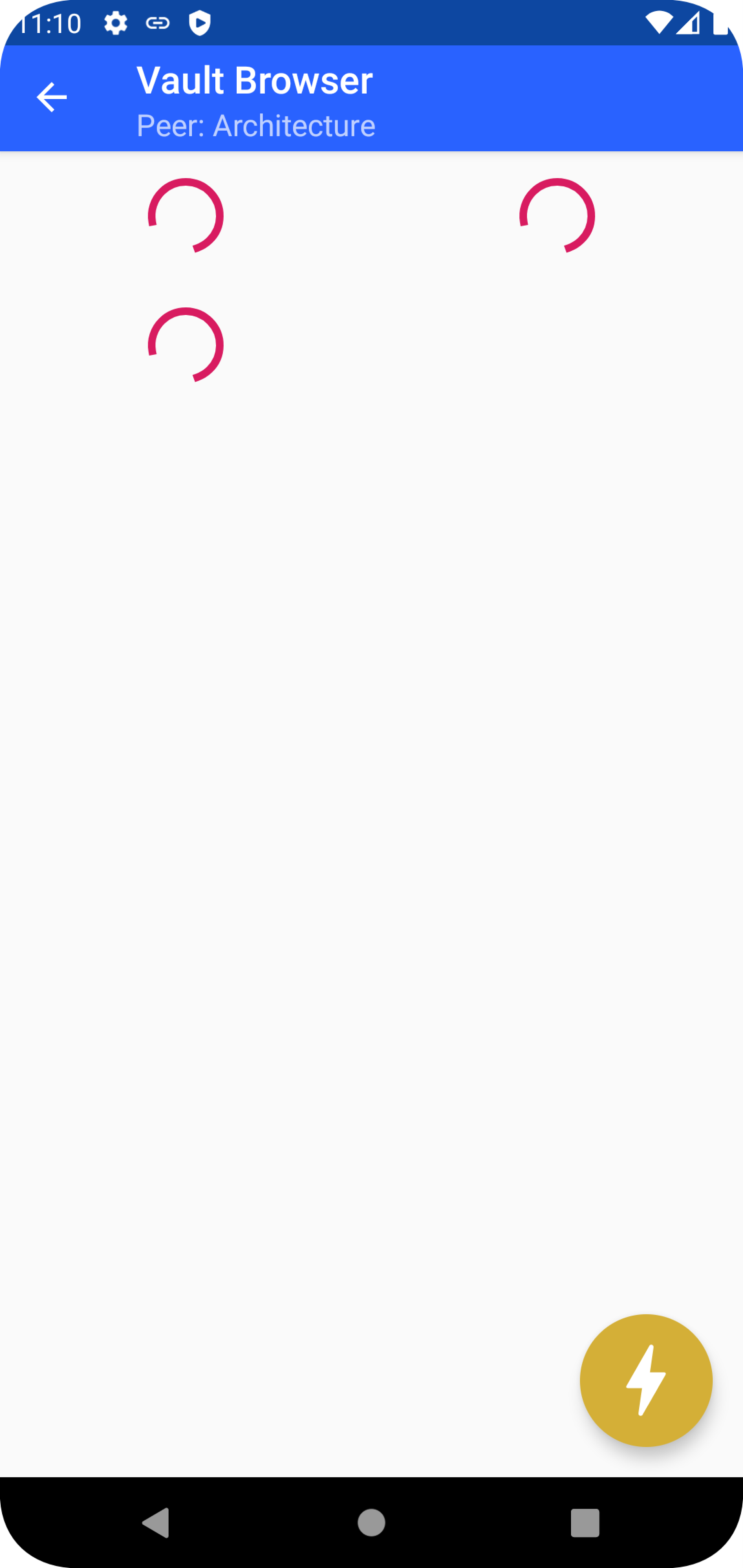}
        \caption{Loading images from a peer.}
        \label{fig:peer-view}
    \end{minipage}
% \end{figure}
\hfill
% \begin{figure}
\begin{minipage}{0.495\columnwidth}
        \centering
        \includegraphics[width=.9\linewidth]{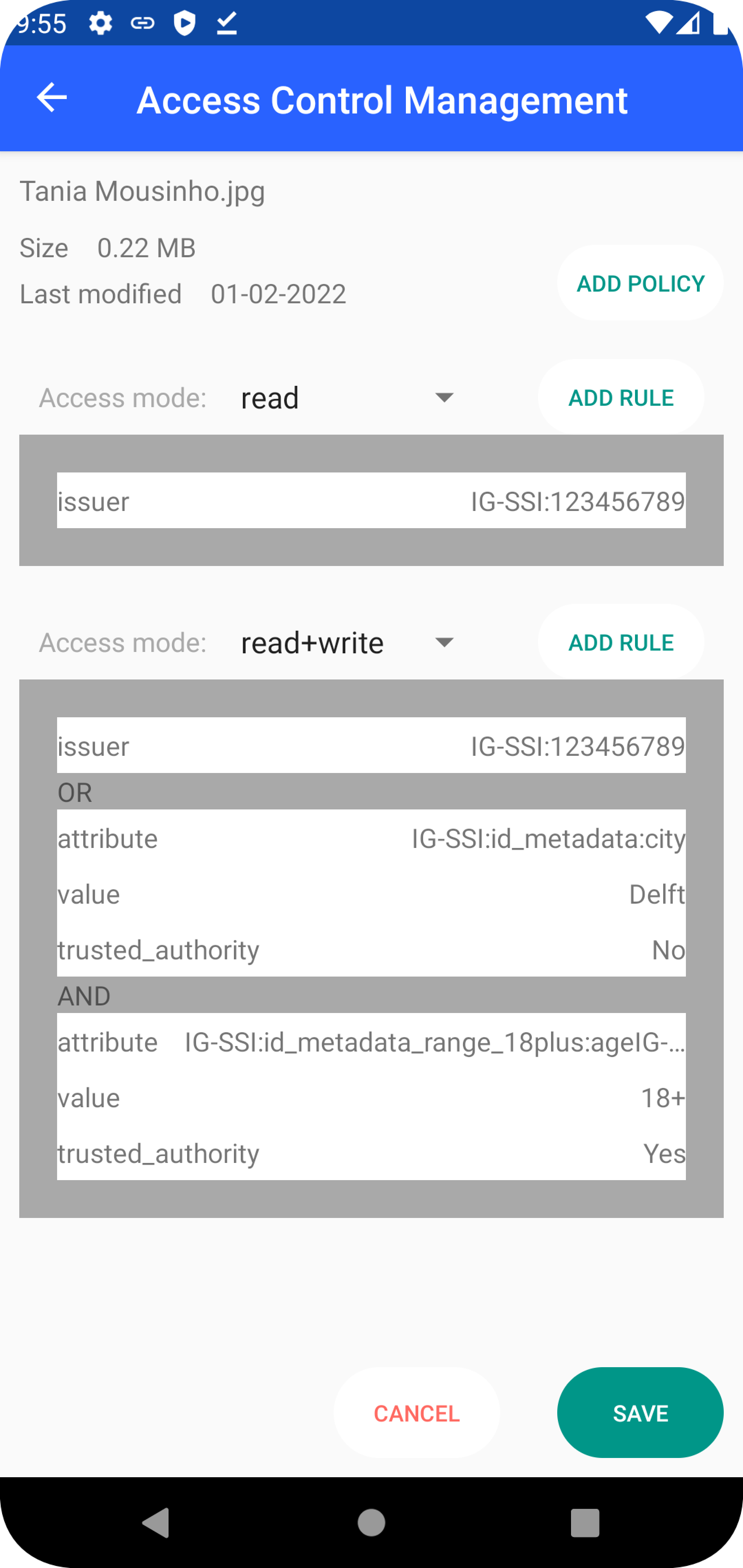}
        \caption{Access Control Management.}
        \label{fig:acm}
    \end{minipage}
    \hfill
    \begin{minipage}{0.495\columnwidth}
        \centering
        \includegraphics[width=.9\linewidth]{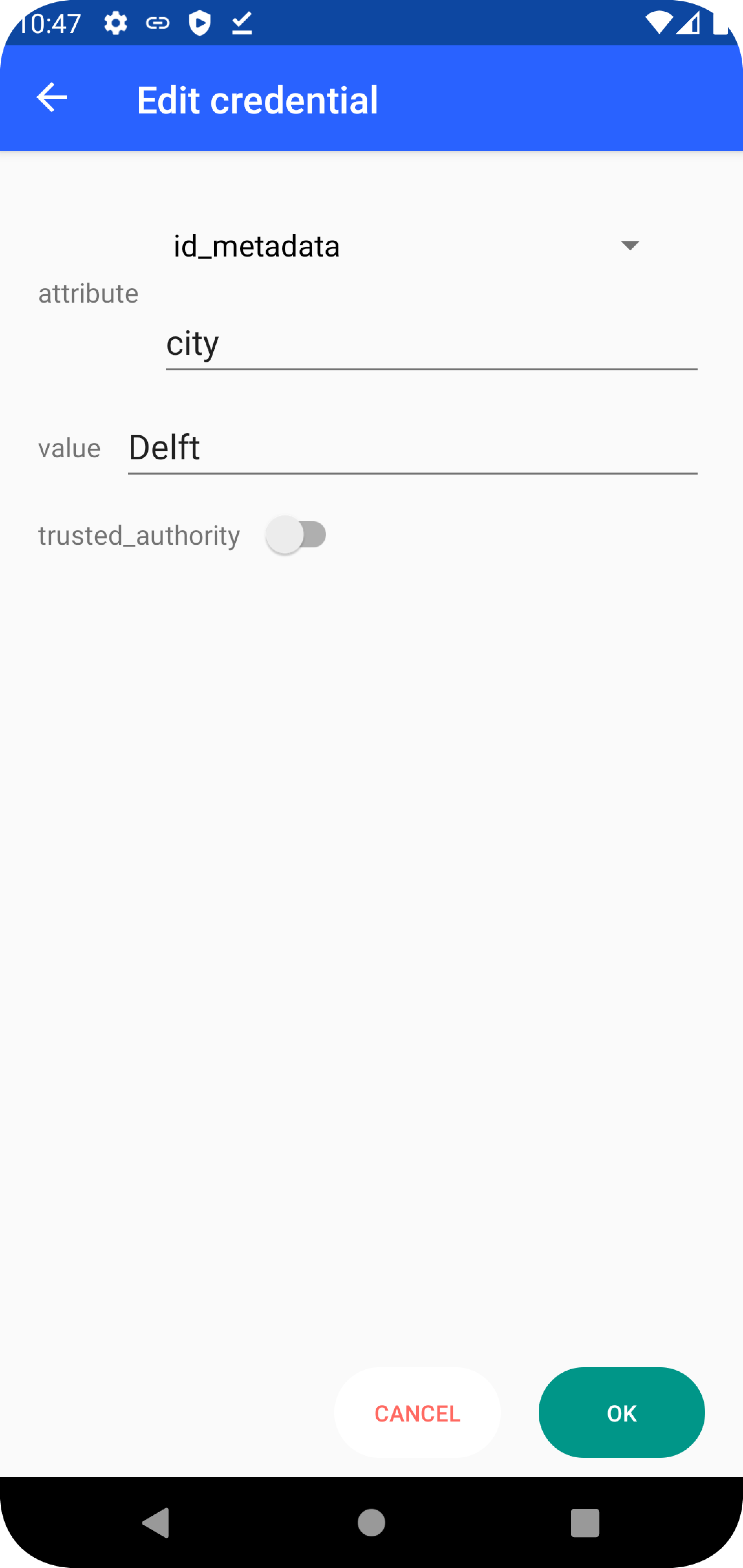}
        \caption{Edit policy rule.}
        \label{fig:edit-credential}
    \end{minipage}
\end{figure*}

Figure \ref{fig:data_vault_components} shows the important components and classes from the data vault. There are three UI views: the VaultBrowser, the AccessControlManagement interface, and the RuleEditor. The VaultBrowser is the main view that lets you explore the photos in your DataVault. The user can select a peer in the DataVaultCommunity to request files from to explore in the VaultBrowser in peer view. The DataVaultCommunity is connected to the TCIDDatabase from which it can retrieve \ac{TCID} attestations, and the EBSIWallet which contains EBSI credentials that along with cached session tokens are used as access tokens in requests to other peers. The TCIDDatabase is managed by a separate Super App application. The DataVaultCommunity runs independently in the background, processing incoming requests. Each message type (e.g. \textit{FileRequestFailed}) has a corresponding message handler (\textit{onFileRequestFailed}). \textit{onAccessibleFilesRequests} trigger an access log transaction to TrustChain. The AccessControlFile for each requested file is retrieved and evaluated against the presented access tokens. In the case of session tokens, the embedded directory tree is searched for the requested file. In the case of credentials, the Policy for the required access mode is evaluated with the presented credentials starting from the root Rule of the rule tree. Binary expression rules have a boolean operator and two sub-rules which are recursively evaluated. The leaves of the tree are unary expression attribute rules that are matched against the attributes in the given credentials. Whenever a leaf rule is evaluated, the credentials that don't satisfy the rule are discarded. If the whole tree is evaluated and the set of candidate credentials is not empty, the policy is satisfied. The AccessControlManagement view is where you create and update access policies for a selected file and the RuleEditor lets you edit a specific rule of a policy. Figures \ref{fig:vault-browser}, \ref{fig:peer-view}, \ref{fig:acm} and \ref{fig:edit-credential} show screenshots of the browser interface for local files and peer browsing, and the access control management interfaces.

Before settling on developing TrustVault, work was done on the Super App's messaging app, implementing features like contact sharing to familiarize IPv8. The biggest challenge was making an intuitive \ac{UI} to edit access policies on a small-screen device. The current \ac{UI} does not reflect the tree-like structure of an access policy. Instead, the linear layout enforces a linear evaluation of access policies. A policy $(A \circ B \circ C \circ D)$ would be evaluated as $(A \circ (B \circ (C \circ D)))$. The shape of policy trees is thus limited to be consistent with what the user expects from the \ac{UI}. 

% \subsection{EBSI data wallet}\label{sec:ebsi-data-wallet}
TrustVault is designed to be a secure data wallet for EBSI users. The process of getting TrustVault EBSI conformant has not been straightforward and is still ongoing. The first prototypes were built using the early versions of the TypeScript cef-ebsi packages to interact with EBSI v1 \citep{npm-cef-ebsi} as part of the EBSI Early Adopters programme\footnote{This work was facilitated and sponsored by The National Office for Identity Data (RvIG) of the Dutch government.}. In v1, all operations were API calls to test endpoints. In v2, critical operations, including creating, signing, and verifying credentials, were moved from the endpoints to libraries running on the user's device. At this point, there were three documentation sources for implementing an EBSI wallet that were out of sync in several places and there was no official library for Android, meaning that there was much trial-and-error to get the API connection working\footnote{Discrepancies in documentation and trial-and-error: https://github.com/Tribler/tribler/issues/6023\#issuecomment-908087676}\footnote{Contact with EBSI support about downtime and errors: https://github.com/Tribler/tribler/issues/6023\#issuecomment-1104821838}. As some wallets started passing the conformance tests, EBSI started publishing test reports that included correctly formulated HTTP requests for the different APIs. We were able to use some of these, including onboarding, authentication and authorisation requests to validate our own implementation up to that point.\\
When initializing TrustVault, the user needs to complete the EBSI onboarding process, which entails scanning a QR-code on the onboarding page to get an authentication token that is used to get permanent authorisation. In subsequent sessions, the authorisation token is exchanged for a short-term session token that needs to be included in every API request.

\subsection{Privacy}
Privacy in TrustVault can be analysed from the perspective of the TrustVault owner and the requesting party.
One of the main goals of this work is to give users control over their data and thus over their privacy. The first step is to enable users to self-host their data, stopping data-hungry companies from running machine learning algorithms over user data and learning users' behavioral patterns. This has the added benefit of disrupting Big Tech from monetizing user data. Giving the user fine-grained access control allows the user to have specific disclosure policies at the desired granularity level, down to the file level. This comes with great responsibility, as there is the opportunity to make mistakes when defining access policies and disclosing data to unintended parties. The challenge is to make the user experience simple and intuitive to minimise the chances of mistakes. Hosts are encouraged to exercise data minimisation: the practice of requesting only the minimum amount of information necessary for an operation. In this context, it means not having policies that require provers to reveal an unnecessary amount of (personal) information. The requesting party, meanwhile, has full control over its identity. Selective disclosurability allows the requesting party to only present information it is comfortable disclosing.\\

Identification by static public keys does present the possibility of learning information over time. The host can keep a record of every time a certain public key wants to access data, which is arguably a sensible thing for the host. However, the host is able to link different access requests over time while collecting the credentials presented at each request, possibly accumulating a more revealing or even identifiable set of attributes of the requesting party. Entities are able to have multiple \acp{DID} for different contexts in \ac{SSI}. This reduces the linkability of credentials to an entity. However, users still have one public key by which they are identified in IPv8 communities, voiding the benefits of having multiple \acp{DID} in EBSI. The Python implementation of IPv8\footnote{https://github.com/Tribler/py-ipv8} has Network-Level Anonymity which mitigates credential linkability and correlation attacks. The Kotlin implementation\footnote{https://github.com/Tribler/kotlin-ipv8} however does not have this feature. TrustChain does not support private/anonymous transactions. By logging access requests on-chain, interactions between parties become publicly visible. Anyone can keep track of when and how often one public key requests access from another public key, potentially leaking information.

\subsection{Security}
The security of TrustVault depends on the security of the data vault and the digital wallet. The data vault's main task is keeping data confidential. The Android internal file storage protects files from being accessed outside the Super App \cite{android-storage}. This offers the first layer of protection. Additionally, the data vault is encrypted using AES when the application is not in use. When opened, a password is required to decrypt the data vault. This prevents unauthorised access even if someone gets physical access to the device and launches the application. Data is also protected in transit with IPv8. Packets are encrypted with the recipient's public key and signed for authenticity and integrity.

TrustVault inherits the \ac{VC} trust model. EBSI can be trusted to be tamper-evident in fulfilling the role of \ac{VDR} by using a public blockchain. It is less convincing in meeting the requirement of accuracy because there is a layer between users and the blockchain where read and write operations could be corrupted. The likeliest way an attacker could get access to data not intended to be disclosed is by getting a false \ac{VC} from a malicious, compromised or incompetent issuer. Issuers that have a reputation to protect are incentivised to be honest. EBSI tries to facilitate this by having an accreditation process for issuers on the \ac{TIR}. Ultimately the verifier decides whom to trust. \ac{TCID} gives more control to the user in this aspect by using personal Trusted Issuer registries. Attestations can also be revoked in \ac{TCID}, resulting in better credential accuracy.

There are several threats to data availability. The first threat is the lack of redundancy. All the user's data is on a mobile device that can temporarily or permanently be out of service for a number of reasons. If the data vault is not backed up on a more reliable medium, the user risks losing data if the device becomes permanently inaccessible. Limited battery capacity and fluctuating internet speed occasional drops in service level can be expected. Communication on the protocol level is more robust. IPv8 maintains network connectivity between peers even with changing IP addresses and firewall protection. While EBSI uses a distributed ledger, interaction with the ledger goes via the hosted Core Services layer. These hosted services can be a single point of failure that can disrupt \ac{VC} verification.

The tamper-proof access log does not provide indisputable proof that a requester retrieved a file. The record only claims that the requester could access a set of files based on the provided \acp{VC}. A malicious host could add files to the bloom filter that are not actually accessible to the requester. This would be difficult for the requester to detect.

\subsection{Experimental analysis}
For TrustVault to be a viable solution for storing and sharing data, file transfer, including the time it takes to verify access tokens, must be as quick as possible. The factors that determine transfer time are Internet connection speed and latency, transfer protocol speed, and access token verification time which is in turn dependent on the available computing power. Internet connection speed sets the upper limit for achievable transfer speeds but will likely not be the bottleneck. The data transfer protocol used is a connection-less data transfer protocol based on the Trivial File Transfer Protocol. The protocol works around the unreliability of \ac{UDP} to create a protocol that can be used for \ac{P2P} data transfer. An average transfer speed of 260kB/s is achieved over WiFi on the same network and 210kB/s over 4G \cite{bambacht-web3}. File size is capped at 250MB to avoid running out of memory when reconstructing packets. The different access token types incur distinct penalties on transfer time because of the different verification methods. Session tokens need to be verified when presented, which requires an \ac{EC} signature verification. Depending on the size of the encoded \emph{accessibleFilesTree}, a session token adds at least 370 bytes to a request. \ac{TCID} attestations also require an \ac{EC} signature verification. However, if multiple attestations are included in a single request, each has to be separately verified. Each attestation also adds at best 1800 bytes to a request. Similar to \ac{TCID}, each EBSI credential in a \acp{VP} requires an \ac{EC} signature verification but an extra verification is to authenticate the holder of the \acp{VC}. Whereas TCID has a local registry of Trusted Issuer public keys, the \ac{DID} of the issuer and the holder has to be resolved with the EBSI DID Registry API. That is an additional overhead of two HTTP requests per \ac{VP}. EBSI access token adds around 200 bytes for the VP metadata and 1300 bytes at a minimum for each included VC to a request. The EBSI VCs in this experiment contain three attribute claims. We therefore chose to use three TCID attestations, each attesting to one claim. These figures are summarised in table \ref{tab:attypes}.

\begin{table}
\centering
\begin{tabular}{ |c|c|c|c| } 
 \hline
  & Verification & Time & Data \\
  \hline
 Session token & 1 & 130 ms & $~$ 300B \\ 
 \hline
 TCID & 3 & 180 ms & 3 $\times$ 1800B  \\ 
 \hline
 EBSI VC & 1 + 1 & 340 ms & 200B + 1300B \\ 
 \hline
\end{tabular}
\caption{Experimental access token verification time and data overhead.}
\label{tab:attypes}
\end{table}

\subsection{Experimental setup}
The experimental setup consists of one phone, a Samsung Galaxy S8 with 4GB RAM, 4x2.3 GHz + 4x1.7 GHz 8-core CPU running Android 9, and a MacBook Pro running a Google Pixel 4 emulator on Android 11. The Galaxy S8 serves as the host TrustVault and the emulator function as requester. All verification and access control is thus done on the Galaxy S8. The devices are connected to the same WiFi network. We ran 7 experiments detailed in table \ref{tab:experiments}. In each experiment, 50 requests and file transfers are made for each access token type and a baseline with no access token. Transfer time is measured from the moment a request is sent to the moment a response is received. A timestamp is included in the request metadata that is returned in the response and the difference is taken from the timestamp at which the response is received. A single file of 220kB is returned in the response. The chosen file is an arbitrary JPEG compressed to approximately 260kB such that a sub-1-second transfer time would indicate the expected 260kB/s throughput of IPv8 is achieved. The first two experiments compare the transfer times over WiFi and 4G. There is one control experiment over WiFi with a file of 1kB. The control is to single out the influence of the request size on transfer time by having a relatively small response. We then simulated different workloads of 50 requests with the time between requests $\Delta$ changed to vary the number of requests per second that the phone needs to handle. The experiments are run with an automated Kotlin script that generates logs that are processed in Python.

\begin{table}
\centering
\begin{tabular}{ |cc c c| } 
 \hline
  \# & Phone & $\Delta$ & file size\\
  \hline
 1 & WiFi & 5 & 220kB\\ 
 \hline
 2 & 4G & 5 & 220kB\\ 
 \hline
 3 & WiFi & 5 & 1kB\\ 
 \hline
 4 & WiFi & 4 & 220kB\\ 
 \hline
 5 & WiFi & 3 & 220kB\\ 
 \hline
 6 & WiFi & 2 & 220kB\\ 
 \hline
 7 & WiFi & 1 & 220kB\\ 
 \hline
\end{tabular}
\caption{Experiment parameters.}
\label{tab:experiments}
\end{table}

\subsection{Experimental results}
\begin{figure*}
    \begin{minipage}{\columnwidth}
        \centering
        \includegraphics[width=\linewidth]{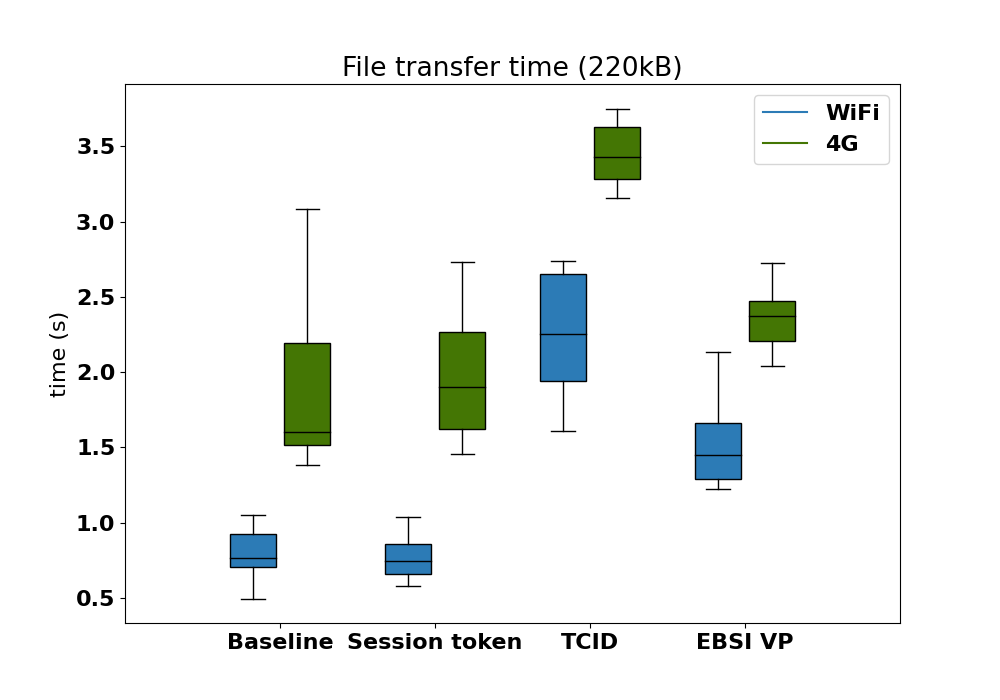}
        \caption{File transfer Time (220kB).}
        \label{fig:file_request_time_single}
    \end{minipage}
    \hfill
    \begin{minipage}{\columnwidth}
        \centering
        \includegraphics[width=\linewidth]{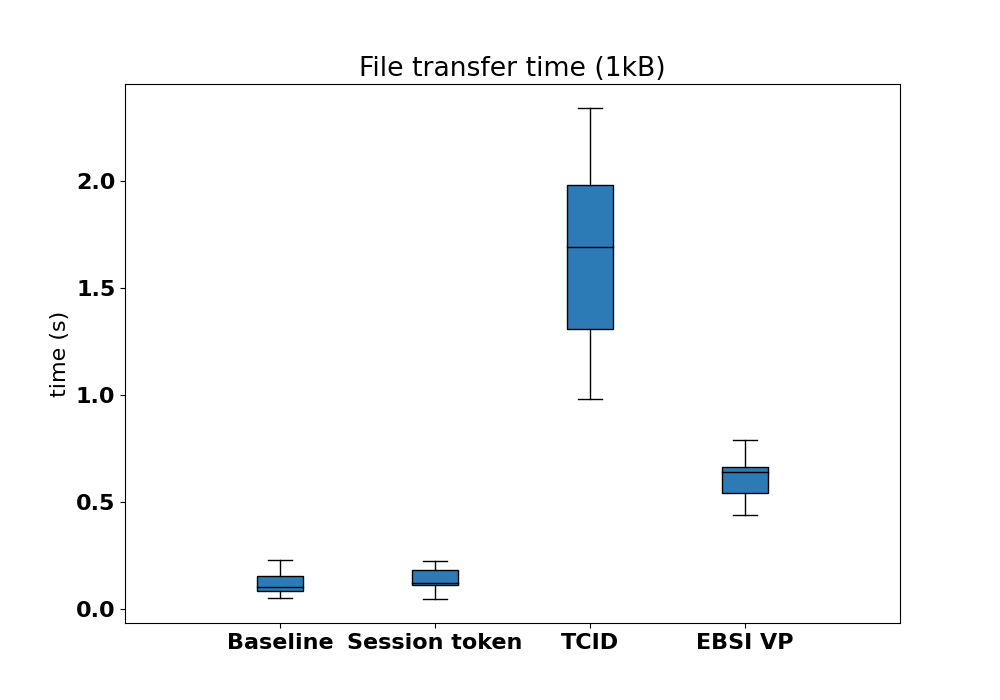}
    \caption{File transfer Time (1kB).}
    \label{fig:file_tranfer_1kb}
    \end{minipage}
\end{figure*}

The WiFi and 4G transfer times are shown in Figure \ref{fig:file_request_time_single}. The results confirm a slower transfer speed when on 4G than on WiFi. The average transfer time increased 138\%, 157\%, 52\% and 54\%, respectively for the four categories. This is a bigger decrease in speed from WiFi to 4G than that measured in \cite{bambacht-web3}. Session token verification is on par with the baseline measurements. The extra signature verification only takes on average 130 ms. It takes about 180 ms to verify the 3 TCID attestations but the bigger request size seems to have the biggest impact on the transfer time. On average, TCID is about 1.5 seconds slower than baseline. EBSI \acp{VP} take the longest to verify at around 340 ms. Yet the overall transfer time is quicker than TCID, and a smaller request size is more optimal. This same pattern holds with the transfer times over 4G. This is confirmed in the results of the control experiment shown in Figure \ref{fig:file_tranfer_1kb}. With a response of just 1kB the transfer time drops with an average of 700 ms. The relative differences between the access token types are preserved.

% \begin{figure}
%     \centering
%     \includegraphics[width=1.0\columnwidth]{images/big-ftt-single-requester.png}
%     \caption{File transfer Time (220kB) 1 requester.}
%     \label{fig:file_request_time_single}
% \end{figure}

\begin{figure}
    \centering
    \includegraphics[width=1.0\columnwidth]{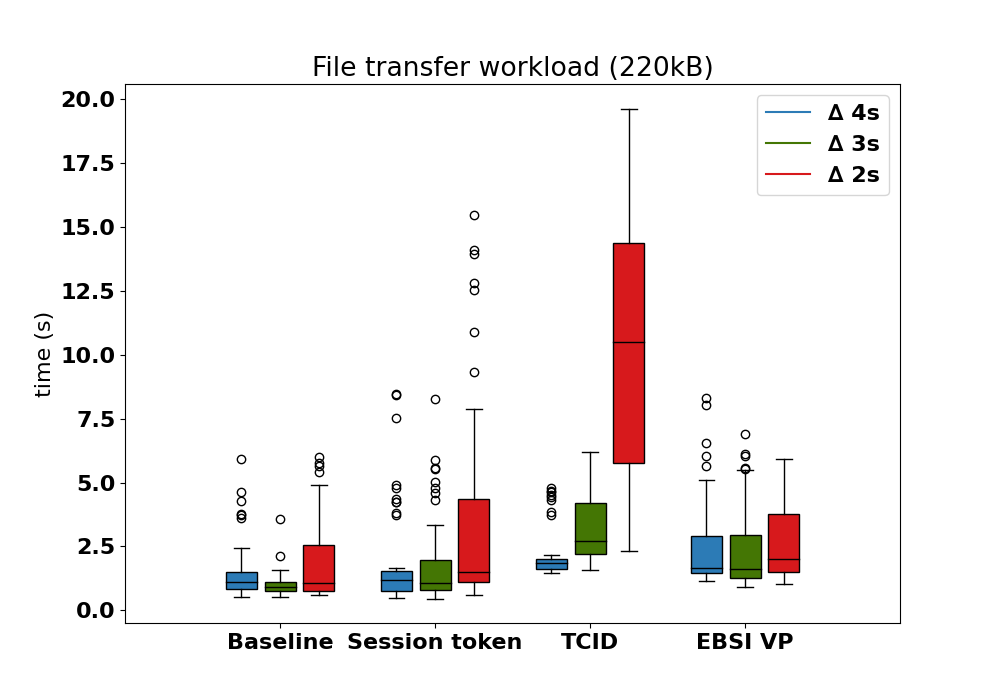}
    \caption{File transfer workload (220kB).}
    \label{fig:file_request_time_double}
\end{figure}

\begin{figure}
    \centering
    \includegraphics[width=1.0\columnwidth]{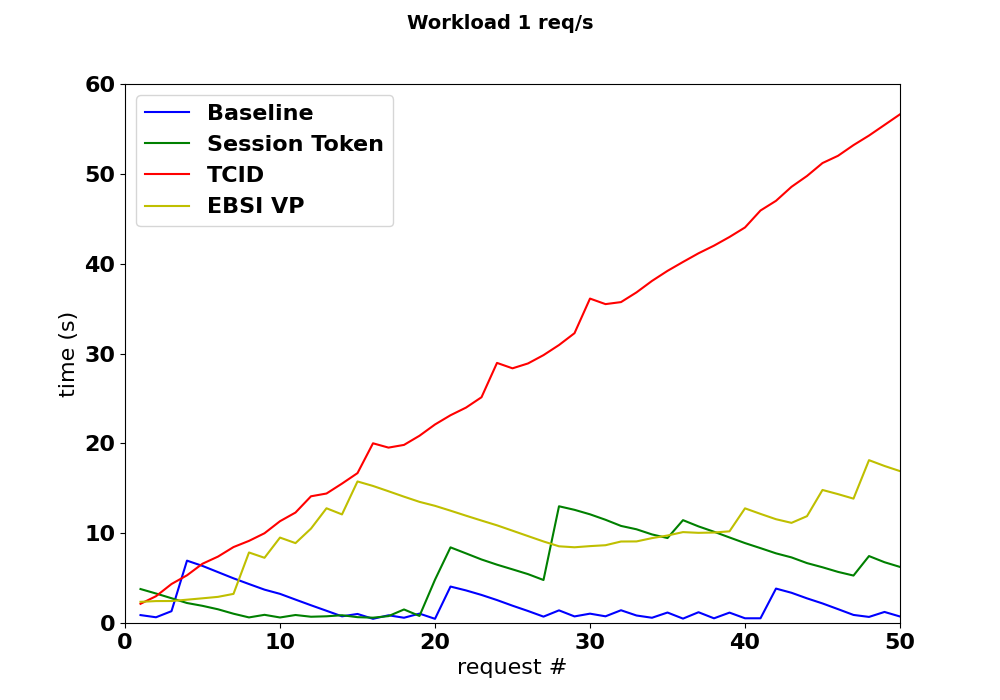}
    \caption{File transfer workload 1 req/s (220kB).}
    \label{fig:workload1}
\end{figure}

The simulated workload results are shown in figure \ref{fig:file_request_time_double}. With a $\Delta=$ 4s, the transfer times mirror that of the initial tests with an interval of 5 seconds. At $\Delta=$ 3s, the baseline, session token, and EBSI VP runs mirror the $\Delta=$ 4s runs but TCID is about 50\% slower. At $\Delta=$ 2s, the phone starts getting overloaded with session tokens and EBSI VPs taking a small hit but TCID taking on average 3.6 times longer. Figure \ref{fig:workload1} shows the transfer times over time for 1 request per second. The CPU load is still low and the phone is still able to verify access tokens nominally, but the network stack gets congested and transfer times gradually increase to tens of seconds in the case of TCID and around 10 seconds for session tokens and EBSI VP.

\subsection{Performance evaluation}
The transfer when the device is not congested is in line with the results of \cite{bambacht-web3}. In the $\Delta=$ 5s baseline case, the mean transfer time for a 220kB file is 780 milliseconds. The request would have a negligible impact on the transfer time as it is just a couple of hundred bytes meaning the file response is the dominant factor. Considering this, the transfer speed can be approximated at around 280kB/s. However, even light congestion causes the transfer time to become impractically high in the order of tens of seconds for a single file. This is a baked-in limitation of IPv8 using UDP. 

On the other hand, the use of access tokens with each request does not diminish the usability of the system, especially when converting TCID and EBSI credentials into session tokens. Session tokens incur almost no time penalty on transfer time. The results show the time penalty incurred for the different types of access tokens on the retrieval of a single file but in practice, multiple files are retrieved with one request, spreading the verification time penalty across multiple files. At the moment, it is fair to say that large-scale data sharing is not yet achieved.
\section{Related Work}\label{sec:related-work}

\newcommand*\rot{\rotatebox{90}}
\newcommand*\OK{\color{green}\ding{51}}
\newcommand*\NOT{\color{red}\ding{53}}

% \makecell{Some really \\ longer text}

\begin{table*}
\centering
\begin{tabular}{|l c c c c c c c c|}
        \hline
          & \rot{\thead{GDPR \\ by design}} & \rot{\thead{Verifiably \\ authenticated \\ AC}} & \rot{\thead{Fine grained \\ AC}} & \rot{\thead{Decentralised}} & \rot{\thead{P2P}} & \rot{\thead{Open source}} & \rot{\thead{Data portability}} & \rot{\thead{Data backup}} \\
        \hline
        \hline
         Solid & \NOT & \NOT & \OK & \OK & \OK & \NOT & \OK & \OK \\
        \hline
         Gaia-X & \NOT & \OK & \OK & \OK & \NOT & \OK & \OK & - \\
        \hline
        Musubi & \OK & \OK? & \NOT & \OK & Cloud relay & \OK & \OK & \NOT \\
        \hline
         Safebook & \NOT & \NOT & \OK & \OK & \OK & \NOT & \NOT & \OK \\
        \hline
         SkyFlow & \NOT & \OK & \OK & \NOT & \NOT & \NOT & \OK & - \\
        \hline
        \cite{kim2021attribute} & \NOT & \OK & \OK & \NOT & \NOT & \NOT & \NOT & - \\
        \hline
        TrustVault & \OK & \OK & \OK & \OK & \OK & \OK & \OK & \NOT \\
        \hline
        \hline
         \cite{rouhani2021distributed} \cite{dam-chain} & - & - & \OK & \OK & \OK & \OK & N/A & N/A \\
        \hline
    \end{tabular}
    \caption{Comparison with related work}
    \label{tab:rw_comparison}
\end{table*}

% \subsection{Solid}
Solid is an open-source protocol that lets people store their data securely in decentralised data stores called Pods \cite{solid-paper}. Pods are personal web servers that can store any kind of data as Linked Data. Linked Data is data with semantic links to other data recorded in its metadata such that computers can explore these links using semantic queries. Pod owners have granular control over who has access to the data. Solid uses the WebAccessControl system, which is based on access control lists with user identification by WebID, to grant and revoke access to any slice of data contained in a Pod to individuals, organizations, or applications. WebID is a protocol that allows persons, organizations or other types of agents to create unique identities and embed links to other people or objects using Resource Description Framework \cite{mainini2016access}. WebID makes it easy to make arbitrary claims about yourself, but those claims are not trustworthy as they are not verifiable. At best, they provide authentication by proving possession over a private key. Access control rules can be based on the agent’s properties in their profile document. Solid applications are client-side, mobile or web, that read data straight from users' pods. Users can switch from one application to the other because data is decoupled from applications by design.

Gaia-X is a federated data infrastructure for data and cloud sovereignty \cite{gaia}. The main problem Gaia-X aims to solve is the lack of trust in the current landscape of storing, sharing and handling data. The Gaia-X infrastructure is based on Ecosystems: an independent group of participants that provide or consume data and services. The infrastructure relies on federated services run by participants to provide trust and data sovereignty in data exchange between participants. These services include decentralised identity management based on \acl{SSI}. Gaia-X employs usage control rather than access control by establishing data contracts between participants and logging data exchanges to enforce the data usage policies in the contracts. Participants have to negotiate a data contract and usage policies for each connection. Usage policies constrain the consumer's use of a resource based on the consumer's properties and verifiable attributes. However, usage policies are not granular and specific to slices of data. While the federation services are not centralised, they publicise information about the relationship between participants and prevent the system from being completely \ac{P2P}. The Gaia-X infrastructure is tailored for business-to-business applications and is not suited for personal and social data sharing.

%Participants can assume the role of Provider, Consumer or both. Providers operate resources and make them available to the ecosystem as a Service Offering. Consumers search for Service Offerings in the ecosystem for consumption. Participants can also become Federators. Federators run Federation Services that sustains the federation and facilitates interaction between participants. Some categories of Federation Services are: Identity \& Trust, Federated Catalogue and Data Sovereignty Services. Identity \& Trust services provides decentralised identity management based on SSI to build trust between participants. Trust is built with Self Descriptions, essentially SSI credentials with information about participants and services. The Federated Catalogue is a resource of Self Description that enables the search for participants, providers and their service offerings in a federation. The Data Sovereignty Services enable users to determine how their data is used by facilitating data contract negotiations between participants and logging when data is submitted and received and if the agreed upon rules were followed.

Musubi \cite{musubi} is a mobile social application for real-time data sharing. It uses the custom Trusted Group Communication Protocol to enable phones to connect through a cloud relay and share data in a group without intermediation. Public keys are used for addressing and communication encryption. Group membership by invitation is the only form of access control.

Safebook \cite{safebook} is another privacy-preserving social application. Safebook uses a P2P overlay with pseudonymous identifiers. Data is not retrieved directly from a peer but relayed through layers of nodes surrounding a peer obfuscating communication. The innermost layer consists of direct contacts of the core and each store an encrypted copy of the core's data providing a backup and fallback when the core is offline. Safebook supports some sort of fine-grained access control but since communication is done pseudonymously, we can conclude that it is not based on user credentials.

SkyFlow PII Data Privacy Vault\footnote{https://www.skyflow.com/product/pii-data-privacy-vault} (DPV) is a zero-trust hosted database service focused on protecting personally identifiable information. Authentication and authorisation are required with each access request. It has row and column-level fine-grained access control based on roles and attributes. As it is intended for internal support teams, it does not leverage the benefits of decentralised identity but uses centralised identity instead.

\cite{kim2021attribute} proposes using an \ac{ABAC} scheme based on \acp{DID}, similar to the scheme proposed in this work. The system is used to control access to a centralised platform and the credentials used are from within the ecosystem. We extend this concept to decentralised data sharing with many independent verifiers and exploit the power of cross-domain and societally relevant credentials.

%The platform is the only verifier in the system and there are three fixed issuers within the system. There is actually no requirement for identity portability or protection from unintentional disclosure because the platform is the sole intended recipient of all credentials.

There is multiple research that looked into decentralizing \ac{ABAC}. \cite{rouhani2021distributed} and \cite{dam-chain} proposes using a blockchain for policy enforcement. The task of making policy decisions is handed over to smart contracts. Access policies and user attributes are stored on smart contracts. When access is requested to a resource, a request is made to a smart contract that makes a decision based on the policies and attributes stored on-chain. The decision is returned to the server enforcing the access policy. This approach makes it possible to have the blockchain serve as a decentralised escrow for digital assets. An on-chain access log is automatically created recording the access policy decision, removing the need to have a separate logging mechanism. A drawback to this approach is that updating policies is costly as that requires write operations on the blockchain. Every access request has to go through a smart contract, introducing some latency.

TrustChain is a Sybil-resistant permissionless blockchain \cite{otte2020trustchain}. Transactions are signed by both parties and blocks are chained together to the previous block of both parties. Each maintains its own chain that is tangled with the chains of parties they transacted with. There is no global chain containing every transaction over which consensus has to be made. Modifications or reordering of blocks on one chain can be detected on the chains of counterparties. This way, consensus is achieved between participants of a transaction instead of on a global level.

\ac{TCID} is an \ac{SSI} system designed with performance and security at the networking layer in mind \cite{TCID}. \ac{TCID} provides the properties of Self-Sovereignty and Credibility, but crucially also Network-level Anonymity. Network-level Anonymity is achieved when source and destination addresses are obfuscated. Without this property, it is possible to carry out correlation attacks on credentials exchanges over time, undermining the data disclosure protections of SSI. \ac{TCID} solves this problem by adding an anonymisation layer on top of the communication layer. The anonymisation layer routes identity-based messages through a multi-hop communication channel of randomly selected peers. Increasing the number of intermediaries improves anonymity but also increases latency. \ac{TCID} supports credentials with \acp{ZKP}, including \ac{ZKP} range proofs.

\cite{rowdy-igssi} extends \ac{TCID} with a distributed revocation mechanism. A gossip protocol is used to propagate revocations through a network. Accepting a revocation is at the verifier's discretion. Verifiers keep their own local registry of Trusted Issuers that inform decisions on both verification and revocation.

% \subsection{Walt.id SSI Kit}
The SSI Kit by walt.id is a Self-Sovereign-Identity open source solution, primarily focused on the EBSI/ESSIF ecosystem\footnote{https://github.com/walt-id/waltid-ssikit}. It provides building blocks for key management, issuing, presenting and verifying credentials, and specific EBSI-related functions. Walt.id developed one of the earliest EBSI conformant wallets.

Table \ref{tab:rw_comparison} shows how the different related works compare. Cloud-based data storage means extra work is necessary to ensure data is secure according to GDPR. Some works have fine-grained access control but do not use verifiable authentication. \cite{musubi} relies on real-world trust and proximity for verifiable authentication. By using only public keys as user identities, fine-grained access control would be strenuous. \cite{kim2021attribute}, like Gaia-X and SkyFlow uses SSI for \acf{ABAC}, but does not veer into decentralised data sharing. Gaia-X does use SSI for \ac{ABAC} in the context of data sharing but relies on federators to achieve sovereign data exchange. SkyFlow and \cite{kim2021attribute} are hosted services and not P2P data-sharing infrastructures. Safebook is the only mobile system with a data backup and fallback mechanism but it compromises on security by having copies of user data on devices of other peers. TrustVault compromises on data backup because storage is limited on mobile phones. Backups are technically possible on a separate storage medium, but fallbacks introduce complexity and security concerns.
\section{Conclusion and Future Work}\label{sec:conclussion-fw}
\begin{comment}
A public blockchain could be used to keep record of every access request made to the TrustVault in a manner that is irrefutable. This further improves the security of TrustVault by making it possible to trace back malicious behavior on an auditable, immutable, publicly verified access log. TrustChain does not support writing arbitrary data to the chain. This functionality would have to be implemented in TrustChain, or a separate ledger could be used solely for this purpose. The storage costs and transaction costs for logging every request on chain may be prohibitive without proper scaling solutions.
\end{comment}

This work presents TrustVault, a system where users are sovereign over both their identity and their data. TrustVault is a unique first operational system that builds upon the upcoming European Digital Identity Wallet. Alternatives for Big Tech is an emerging topic of research. In our system, users are not reliant on Big Tech companies to authenticate themselves or store and host their data. User data is stored in a data vault on a device under the control of the user. To our knowledge, TrustVault is the first solution to use \acl{ABAC} to achieve fine-grained access control to data in a peer-2-peer context while leveraging  \ac{SSI} for the wealth of verifiable attributes available. We show that the \ac{EU}'s \ac{EBSI} initiative is a viable way to give control to the citizens of the EU by integrating it into our system in a societally relevant way. Compared to related works Solid and Gaia-X, there is no infrastructure and system management burden for the user. The user does not need to understand the inner workings of data management. Our work presents the first data and identity wallet solution with true sovereignty over both. It is possible to have a fairer and more competitive system than the for-profit infrastructure of Big Tech, which is public, transparent, and open source.\\

\subsection*{Future Work}
TrustVault can be expanded to support other SSI networks like Sovrin and many built on Ethereum. This would open the door to even more types of credentials and attributes to include in access policies. \ac{TCID} supports some \acp{ZKP} but there are currently more proof schemes in development like BBS+ signatures\footnote{https://www.evernym.com/blog/bbs-verifiable-credentials/} that provide selective disclosure, signature blinding and private holder blinding. These schemes further improve user privacy. Network-Level anonymity, which is already implemented in Python, could be implemented in Kotlin as well. This would mitigate the correlation attacks possible in the system as is. Improving the \ac{UI} to reflect the structure of access policies better could allow the user to set up more complex and expressive policies intuitively. For critical data with high availability requirements, having a fallback device could be a great capability. Redundant devices could be deployed simultaneously for load balancing or simply as a backup. Finally, applications can be developed that make use of the TrustVault infrastructure to provide valuable services to TrustChain Super App users.

\bibliographystyle{IEEEtran}
\bibliography{myreferences}

\end{document}